\documentclass[12pt,aps,prd,superscriptaddress,showpacs,longbibliography,floatfix,nofootinbib]{revtex4-1}

\usepackage[utf8]{inputenc}
\usepackage{slashed}
\pdfoutput=1

\usepackage{color}
\usepackage{graphicx}   
\usepackage{bm}
\usepackage{amsmath}
\usepackage{amsfonts}
\usepackage{eufrak}
\usepackage{hyperref}

\def\sumint{\,\hbox{$\sum$}\!\!\!\!\!\!\!\int}

\begin{document}

\title{Thermal free energy of large Nf QED in 2+1 dimensions from weak to strong coupling}

\author{Paul Romatschke}
\affiliation{Department of Physics \& CTQM , University of Colorado, Boulder, USA}
\author{Matias S\"appi}
\affiliation{Helsinki Institute of Physics and Department of Physics, University of Helsinki, Finland}

\begin{abstract}
  In 2+1 dimensions, QED becomes  exactly solvable for all values of the fermion charge $e$ in the limit of many fermions $N_f\gg 1$. We present results for the free energy density at finite temperature $T$ to next-to-leading-order in large $N_f$. In the naive large $N_f$ limit, we uncover an apparently UV-divergent contribution to the vacuum energy at order ${\cal O}(\frac{e^6 N_f^3}{\epsilon})$, which we argue to become a finite contribution of order ${\cal O}(e^6 N_f^4)$ when resumming formally higher-order $1/N_f$ contributions. We find the finite-temperature free energy to be well-behaved for all values of the dimensionless coupling $e^2N_f/T$, and to be bounded by the free energy of $N_f$ free fermions and non-interacting QED3, respectively. We invite follow-up studies from finite-temperature lattice gauge theory at large but fixed $N_f$ to test our results in the regime \hbox{$e^2N_f/T\gg 1$}.
\end{abstract}

\maketitle

\section{Introduction}

The conjectured duality between strongly coupled gauge theories and classical gravity in one higher dimension has been an extremely successful tool to effectively calculate properties of large $N$ gauge theories at strong coupling and finite temperature \cite{Maldacena:1997re,Gubser:1998nz,Itzhaki:1998dd,Policastro:2001yc}.

Unfortunately, while generally expected to be correct, there is no formal proof of the conjecture. Furthermore, only certain gauge theories have known gravity duals, and this list does not include gauge theories that are realized in nature such as QED or QCD. Finally, while gauge-gravity duality allows calculations in a regime where the coupling of the field theory is effectively infinite, the gravity dual is just as hard (or harder) to solve than the original field theory for intermediate values of the coupling, which are often physically relevant.

This provides the motivation to revisit and generalize existing tools to solve quantum field theories (and specifically gauge theories realized in nature) at finite temperature for arbitrary (weak or strong) values of the coupling. At first glance, this project seems to be dead on arrival: if techniques existed to, say, solve QCD non-perturbatively, using gauge-gravity dual results for ${\cal N}=4$ super-Yang--Mills theory as a proxy for QCD would not have been needed. Surprisingly, however, a number of large $N$ quantum field theories can be solved at finite temperature for all values of the coupling, including scalar field theories \cite{Drummond:1997cw,Romatschke:2019ybu,DeWolfe:2019etx}, Wess--Zumino models \cite{DeWolfe:2019etx} and Gross--Neveu models, albeit in two spatial dimensions (2+1d).

In 3+1 dimensions, divergences requiring a renormalization program spoil much of the beauty of the exact (and sometimes analytic) results found in 2+1d. This typically leads to the large $N$ 3+1-dimensional theories exhibiting a Landau pole, as is the case for scalar theories \cite{Romatschke:2019gck} and four-dimensional QED \cite{Moore:2002md,Ipp:2003zr}. While the theories are still useful in the effective theory sense, cut-off effects near the Landau pole imply that in 3+1 dimensions, the strong-coupling limit of these theories is ambiguous.

For this reason, we are led to consider QED in 2+1 dimensions (``QED3'') at finite temperature in the limit of many fermions $N_f\gg 1$, which is free of a Landau pole, and hence is unambiguously defined for any value of the coupling (cf. Refs.~\cite{DHoker:1981bjo,Pisarski:1984dj,Appelquist:1988sr}). Because the theory does not exhibit any logarithmic divergences at leading and next-to-leading order in large $N_f$, in the massless fermion case QED3 is essentially a finite quantum field theory, and there are no logarithmic scale dependencies in the coupling. This implies that the free energy $f\propto T^3$ of QED3 scales as the third power of the temperature, with a coefficient that is only dependent on the (dimensionless) coupling $\frac{e^2 N_f}{T}$. 

In this work, we determine the ratio $f/T^3$ in QED3 non-perturbatively for all (weak to strong) values of the dimensionless coupling $\frac{e^2 N_f}{T}$ to NLO at large $N_f$ using well-established field theory techniques. Our results thus generalize studies of QED3 at $T=0$ \cite{Pisarski:1984dj,Appelquist:1988sr} to arbitrary temperature, and may be useful as a reference for lattice gauge theory studies \cite{Azcoiti:1993ey,Lee:2002id,Hands:2002dv,Strouthos:2008kc,Raviv:2014xna,Karthik:2015sgq}, dualities found for ``cousins'' of QED in 2+1 dimensions  \cite{Son:2015xqa,Karch:2016sxi,Karch:2016aux}, conformal QED3 studies  \cite{Kaul:2008xw,Giombi:2016fct}, as well as condensed matter systems ~\cite{Franz:2002qy}.

\section{Setup}

Let us consider QED with $N_f$ massless fermions defined by the Lagrangian
\begin{equation}
  \label{eq:l1}
  {\cal L}=-\frac{1}{4}F_{\mu\nu} F^{\mu\nu}+\bar\psi_a \left(i\slashed{\partial}+e \slashed{A}\right)\psi_a\,,
  \end{equation}
where $A_\mu$ is the photon gauge field, $F_{\mu\nu}=\partial_\mu A_\nu-\partial_\nu A_\mu$ is the photon field strength tensor, $\psi_a$ with $a=1,2,\ldots,N_f$ are four-component spinors (both in $D=3$ and $D=4$ space-time dimensions), $e$ is the fermion charge, and $\slashed{A}=\gamma_\mu A^\mu$. The Lagrangian (\ref{eq:l1}) is manifestly invariant under gauge transformations. QED at finite temperature $T$ may be defined as given by the Lagrangian (\ref{eq:l1}) with imaginary time on a Euclidean manifold, with the time-like direction compactified on a circle with radius $\beta=T^{-1}$ (see e.g. \cite{Laine:2016hma}). The resulting $D$-dimensional Euclidean action is given by
\begin{equation}
  S_E=\int d^Dx \left[\frac{1}{4}F_{\mu\nu}F_{\mu\nu}+\bar\psi_a\left(\slashed{\partial}-i e \slashed{A}\right)\psi_a\right]\,,
  \end{equation}
where $A_\mu,\psi_a$ are the Euclidean versions of the gauge field and the fermion, respectively, and $\slashed{A}=\gamma_\mu^E A_\mu$ with $\gamma_0^E=\gamma^0,\gamma_{1,2,3}^E=-i \gamma^{1,2,3}$ the Euclidean $\gamma$-matrices. Note that while the gauge field obeys periodic boundary conditions in the time-like direction, the fermions require anti-periodic boundary conditions.

Gauge invariance of $S_E$ implies that there are gauge configurations $A_\mu$ along which $S_E$ does not change. The existence of these ``flat directions''
implies that the QED partition function, defined as $Z=\int {\cal D}A e^{-S_E}$, is ill-defined, because integration along the flat directions leads to divergences\footnote{Note that this is different when choosing a compact formulation of the Lagrangian by trading the gauge field $A_\mu$ with a compact link variable $U=e^{i A_\mu}$.}. In order to make sense of the theory in the non-compact formulation, it is necessary to break gauge invariance. This is customarily done using the Faddeev--Popov formalism by introducing the ghost fields $\bar{c},c$, such that for instance in the class of covariant gauges the gauge-fixed Euclidean action becomes \cite{Laine:2016hma}
\begin{equation}
  \label{eq:segf}
  S_E=\int d^Dx \left[\frac{1}{4}F_{\mu\nu}F_{\mu\nu}+\bar\psi_a\left(\slashed{\partial}-i e \slashed{A}\right)\psi_a+\frac{1}{2\xi}\left(\partial_\mu A_\mu\right)^2+\partial_\mu \bar c\partial_\mu c \right]\,,
  \end{equation}
where the anti-commuting ghosts fulfill periodic boundary conditions just like the bosonic gauge field. The partition function defined from the gauge-fixed action (\ref{eq:segf}) is well-defined, and hence (\ref{eq:segf}) will be used as the definition of QED in the following. While not gauge invariant, the action (\ref{eq:segf}) is invariant under BRST transformations
\begin{eqnarray}
  \delta A_\mu=\partial_\mu c \zeta\,,\quad
  \delta \bar c=\frac{1}{\xi} \partial_\mu A_\mu \zeta \,,\quad
  \delta c=0\,,\quad
  \delta \bar\psi_a=-i e c \bar \psi_a \zeta \,,\quad
  \delta \psi_a=i e c \psi_a \zeta\,,
\end{eqnarray}
where $\zeta$ is an anti-commuting space-time independent parameter such that $\{\zeta,c\}=\{\zeta,\bar c\}=\{\zeta,\psi_a\}=\{\zeta,\bar \psi_a\}=0$. BRST invariance of the action (\ref{eq:segf}) guarantees that many important features of gauge theories, such as Ward--Takahashi identities, are maintained even if gauge invariance has been broken.

The gauge-fixed Euclidean action (\ref{eq:segf}) may be used to evaluate properties of QED at finite temperature perturbatively when expanding $e^{-S_E}$ in a Taylor series around vanishing coupling $e=0$. However, it is possible to resum an infinite number of contributions in this Taylor series by suitably rewriting $S_E=S_0+S_I$, for instance with
\begin{eqnarray}
  \label{eq:eff}
  S_0&=&\int d^Dx \left[\frac{1}{4}F_{\mu\nu}F_{\mu\nu}+\bar\psi_a\left(\slashed{\partial}+m\right)\psi_a+\frac{1}{2\xi}\left(\partial_\mu A_\mu\right)^2+\partial_\mu \bar c\partial_\mu c \right]+\frac{1}{2}\int d^D x d^Dy A_\mu\Pi_{\mu\nu}A_\nu\,,\nonumber\\
  S_I&=&- i e \int d^Dx \bar\psi_a\slashed{A}\psi_a-\frac{1}{2}\int d^D x d^Dy A_\mu(x)\Pi_{\mu\nu}(x-y)A_\nu(y)\,,
\end{eqnarray}
where the same term was added and subtracted in (\ref{eq:segf}). Using $S_0$ instead of (\ref{eq:segf}) with $e=0$ as the reference action allows one to non-perturbatively resum an infinite number of Feynman diagrams (``Dyson series''). Nevertheless, it is important to maintain BRST invariance of $S_0$ in order to avoid introducing gauge-dependent artifacts. One finds that BRST invariance of $S_0$ requires $\partial_\mu \Pi_{\mu\nu}=0$, which is a condition that we will check \textit{a posteriori}.

\subsection*{Photon Self-Energy}

As in Refs.~\cite{Romatschke:2019rjk,Romatschke:2019wxc}, the quantity $\Pi_{\mu\nu}$ is fixed by calculating the full connected photon two-point function, which in the limit $N_f\rightarrow \infty$ becomes
\begin{eqnarray}
  G_{\mu\nu}(x)&=&\langle A_\mu(x)A_\nu(0)\rangle\,,\\
  &=&G_{\mu\nu}(x)+\int_{y,z} G_{\mu\alpha}(x-y)\left(\Pi_{\alpha\beta}(y-z)-e^2 \langle\bar\psi_a(y)\gamma_\alpha\psi_a(y)\bar\psi_b(z)\gamma_\beta \psi_b(z)\rangle\right)G_{\beta\nu}(z)\nonumber\,,
\end{eqnarray}
or, taking into account the extra minus sign arising from the fermion loop,
\begin{equation}
  \Pi_{\mu\nu}(x)=-e^2 N_f {\rm Tr}\left(\Delta(x)\gamma_\mu \Delta(-x)\gamma_\nu\right)+{\cal O}(N_f^0)\,,\quad \Delta(x)=\frac{1}{N_f}\langle \bar \psi_i(x)\psi_i(0)\rangle\,.
  \end{equation}
Note that here $G_{\mu\nu}(x),\Delta(x)$ denote fully dressed propagators, but to leading order in large $N_f$ we can take the fermion propagator $\Delta(x)$ to be free. It is easiest to express the $\Delta(x)$ by going to Fourier space where
\begin{equation}
  \psi(x)=\sumint_{\{K\}} e^{i K\cdot x}\psi(K)\,,\quad \sumint_{\{K\}}\equiv T \sum_{\{\omega_n\}} \mu^{2 \epsilon}\int \frac{d^{D-1}k}{(2 \pi)^{D-1}}\,,
\end{equation}
where $\{\omega_n\}=\pi T (2n +1)$ with $n\in \mathbb{Z}$ are the fermionic Matsubara frequencies, $\mu$ is the renormalization scale parameter and we use dimensional regularization with $\epsilon>0$. With these conventions, the free fermion propagator becomes
\begin{equation}
  \Delta(x)=\sumint_{\{K\}} \frac{e^{-i K\cdot x}\left(-i \slashed{K}\right)}{K^2}\,,
  \end{equation}
which leads to the photon self-energy given by
\begin{equation}
  \Pi_{\mu\nu}(x)=-e^2 N_f \sumint_{\{K\},\{Q\}}e^{-i (K-Q)\cdot x} \frac{{\rm Tr}\left[(-i \slashed{K})\gamma_\mu(-i \slashed{Q})\gamma_\nu\right]}{K^2Q^2}\,.
  \end{equation}
The trace is readily evaluated using the properties of $\gamma$-matrices, finding
$${\rm Tr}\left[(-i \slashed{K})\gamma_\mu(-i \slashed{Q})\gamma_\nu\right] = 4 \delta_{\mu\nu}K\cdot Q-4 K_\mu Q_\nu-4K_\nu Q_\mu\,.$$
In Fourier space, the photon self-energy thus becomes
\begin{equation}
  \label{eq:matterpi}
  \Pi_{\mu\nu}(P)=-4 e^2 N_f \sumint_{\{K\}}\frac{\delta_{\mu\nu}\left(K^2-P\cdot K\right)-2 K_\mu K_\nu+K_\mu P_\nu+K_\nu P_\mu}{K^2(K-P)^2}\,.
\end{equation}

Let us first calculate the zero-temperature (vacuum) part of $\Pi$, which is given by
\begin{equation}
  \Pi_{\mu\nu}^{T=0}(P)=-4 e^2 N_f \mu^{2 \epsilon}\int \frac{d^{D}K}{(2 \pi)^D}\int_0^1 dx\frac{\delta_{\mu\nu}\left(K^2-P\cdot K\right)-2 K_\mu K_\nu+K_\mu P_\nu+K_\nu P_\mu}{[K^2x+(K-P)^2(1-x)]^2}\,.
\end{equation}
Shifting the integration variable $K\rightarrow K+(1-x)P$, the momentum integration is straightforward in dimensional regularization where $D=3\rightarrow 3-2\epsilon$ with $\epsilon>0$.
One finds
\begin{equation}
  \label{eqpiv11}
\lim_{m\rightarrow 0}  \Pi_{\mu\nu}^{T=0}(P)=\frac{8 e^2 N_f}{(4 \pi)^{D/2}} \mu^{2 \epsilon}\left(\delta_{\mu\nu}-\frac{P_\mu P_\nu}{P^2}\right) \Gamma\left(2-\frac{D}{2}\right)\frac{\Gamma^2\left(\frac{D}{2}\right)}{\Gamma(D)} (P^2)^{D/2-1}\,.
\end{equation}

There are no logarithmic divergences in dimensional regularization, and one can take the limit $\epsilon\rightarrow 0$, finding \cite{Pisarski:1984dj}
\begin{equation}
  \label{eq:pivacd3}
  \Pi_{\mu\nu}^{T=0,D=3}(P)=\frac{e^2 N_f}{8} \left(\delta_{\mu\nu}-\frac{P_\mu P_\nu}{P^2}\right) \sqrt{P^2}\,.
\end{equation}


At finite temperature, Lorentz covariance is broken through the presence of a local matter rest frame. This implies that $\Pi_{\mu\nu}$ may be decomposed into the most general tensor structure that can be built out of $\delta_{\mu\nu},P_\mu$ and the rest frame vector $n_\mu=\left(1,{\bf 0}\right)$. The corresponding decomposition is standard in quantum field theory (cf. Ref.\cite{Kraemmer:2003gd}) and we use the complete and orthogonal tensor basis spanned by
\begin{equation}
  \label{eq:projectors}
  {\cal A}_{\mu\nu}\equiv \delta_{\mu\nu}-\frac{P_\mu P_\nu}{P^2}-\frac{\tilde n_\mu \tilde n_\nu}{\tilde n^2}\,,\quad
  {\cal B}_{\mu\nu}\equiv \frac{\tilde n_\mu \tilde n_\nu}{\tilde n^2}\,,\quad
  {\cal C}_{\mu\nu}\equiv \frac{P_\mu P_\nu}{P^2}\,,\quad
  {\cal D}_{\mu\nu}\equiv \tilde n_\mu P_\nu+\tilde n_\nu P_\mu\,,
\end{equation}
to evaluate the structure functions for $\Pi_{\mu\nu}=\Pi_A {\cal A}_{\mu\nu}+\Pi_B {\cal B}_{\mu\nu}+\Pi_C {\cal C}_{\mu\nu}+ \Pi_D {\cal D}_{\mu\nu}$. Here $\tilde n_\mu\equiv n_\nu \left({\cal A}_{\mu\nu}+{\cal B}_{\mu\nu}\right)$. Evaluating $P_\mu \Pi_{\mu\nu}$ from (\ref{eq:matterpi}) one finds
\begin{equation}
  P_\mu \Pi_{\mu\nu}(P)=-4 e^2 N_f \sumint_{\{K\}}\frac{P_\nu K^2+K_\nu \left(P-K\right)^2-K^2 K_\nu }{K^2(K-P)^2}=0=\Pi_C P_\nu+P^2\Pi_D\tilde n_\nu\,,
  \end{equation}
which implies $\Pi_C=\Pi_D=0$ and confirms that BRST invariance is satisfied for the action (\ref{eq:eff}). The structure functions $\Pi_A,\Pi_B$ may be found by considering the components
\begin{eqnarray}
  \Pi_{\mu\mu}&=&(D-2)\Pi_A(P)+\Pi_B(P)=-4 (D-2) e^2 N_f \sumint_{\{K\}}\frac{K^2-P\cdot K}{K^2(K-P)^2}\,,\\
  \Pi_{00}&=&\frac{|{\bf p}|^2}{P^2}\Pi_B(P)=-4 e^2 N_f \sumint_{\{K\}}\frac{K^2-P\cdot K-2 k_0^2+2 k_0 p_0}{K^2(K-P)^2}\,.
\end{eqnarray}
The corresponding thermal sums may be evaluated using standard finite-temperature field theory methods \cite{lebellac_1996}, and the finite temperature parts are given for instance in the appendix of Ref.~\cite{Carrington:2019ggz}:
\begin{eqnarray}
  \Pi_{\mu\mu}^{T\neq 0}=4 (D-2) e^2 N_f {\rm Re} \int\frac{d^{D-1}{\bf k}}{(2\pi)^{D-1}}\frac{n_F(|{\bf k}|)}{|{\bf k}|} \frac{2 i p_0 |{\bf k}|+2 {\bf p}\cdot {\bf k}}{2 i p_0 |{\bf k}|+2 {\bf p}\cdot{\bf k}-P^2}\,,\nonumber\\
  \Pi_{00}^{T\neq 0}=4 (D-2) e^2 N_f {\rm Re} \int\frac{d^{D-1}{\bf k}}{(2\pi)^{D-1}}\frac{n_F(|{\bf k}|)}{|{\bf k}|} \frac{ i p_0|{\bf k}|+2 {\bf k}^2-{\bf p}\cdot {\bf k}}{2 i p_0 |{\bf k}|+2 {\bf p}\cdot{\bf k}-P^2}\,,
  \end{eqnarray}
where ${\rm Re}f(p_0)=\frac{1}{2}\left(f(p_0)+f(-p_0)\right)$. For $D=3$, the remaining angular integration may be carried out to find
\begin{eqnarray}
  \label{eq:matterpis}
  \Pi_{\mu\mu}^{T\neq 0,D=3}=4 e^2 N_f {\rm Re} \int_0^\infty \frac{d k\,  n_F(k) }{2 \pi} \left(1-\frac{\sqrt{P^2}}{\sqrt{P^2-4 k^2-4 i p_0 k}}\right)\,,\nonumber\\
  \Pi_{00}^{T\neq 0,D=3}=4 e^2 N_f {\rm Re}\int_0^\infty \frac{d k\,  n_F(k) }{2 \pi} \left(1-\frac{\sqrt{P^2-4 k^2-4 i p_0 k}}{\sqrt{P^2}}\right) \,.
  \end{eqnarray}

\section{Partition Function for QED3}

The partition function for QED3 is given by
\begin{equation}
  Z=\int {\cal D}\bar\psi {\cal D}\psi {\cal D}\bar c{\cal D}c {\cal D}A e^{-S_0-S_I}\,,
  \end{equation}
with $S_0,S_I$ given in Eqns.~(\ref{eq:eff}). To leading and next-to-leading order in large $N_f$, $S_0$ already resums all the relevant ``daisy-type'' diagram contributions, such that contributions from $S_I$ only appear at order ${\cal O}(N_f^{-1})$, which we neglect. Hence the free energy density for QED3 to NLO in large $N_f$ is given by
\begin{equation}
  f=-\frac{T}{V}\ln Z=f_{\rm ghost}+f_{\rm fermion}+f_{\rm photon}\,,
\end{equation}
with
\begin{eqnarray}
  f_{\rm ghost}=-\sumint_{K}\ln K^2\,,\quad
  f_{\rm fermion}=-2 N_f\sumint_{\{K\}} \ln K^2\,,\quad
  f_{\rm photon}=\frac{1}{2}\sumint_K \ln {\rm det}\, G^{-1}_{\mu\nu}(K)
  \end{eqnarray}
where we used the fact that all the path integrals are Gaussian in momentum space. Here $G^{-1}_{\mu\nu}(K)$ is the inverse photon propagator in momentum space, which from the expression given in $S_0$ takes the form
\begin{eqnarray}
  G_{\mu\nu}^{-1}&=&K^2 \delta_{\mu\nu}-K_\mu K_\nu\left(1-\frac{1}{\xi}\right)+\Pi_{\mu\nu}(K)\,,\nonumber\\
  &=&\left(K^2+\Pi_A(K)\right){\cal A}_{\mu\nu}+\left(K^2+\Pi_B(K)\right){\cal B}_{\mu\nu}+\frac{K^2}{\xi}{\cal C}_{\mu\nu}\,,
\end{eqnarray}
using the projectors given in (\ref{eq:projectors}). The determinant in $f_{\rm photon}$ is given by the product of the eigenvalues of $G_{\mu\nu}^{-1}$, which are the factors multiplying the orthogonal projectors above. Therefore, the ${\cal O}\left(N_f^0\right)$ contribution to the free energy is given by
\begin{equation}
f_{\rm ghost}+f_{\rm photon}=\frac{1}{2}\sumint \ln \left[\frac{\left(K^2+\Pi_A\right)\left(K^2+\Pi_B\right)}{K^2}\right]\,,
\end{equation}
where we used that $\sumint\, \ln \xi=0$ in dimensional regularization. The photon polarization contributions $\Pi_{A,B}$ consist of a zero-temperature piece and a finite-temperature contribution given in (\ref{eq:pivacd3}), (\ref{eq:matterpis}) above, which for $D=3$ become
\begin{eqnarray}
  \label{piexp}
  \Pi_A(P)&=&\Pi_V(P)-\frac{4 e^2 N_f T}{|{\bf p}|^2} \int_0^\infty \frac{dk}{2\pi} n_F(k T)\left(p_0^2+\sqrt{P^2}{\rm Re}\frac{(i p_0+2 k T)^2}{\sqrt{P^2-4 k^2 T^2-4 i p_0 k T}}\right)\nonumber\\
  \Pi_B(P)&=&\Pi_V(P)+4 e^2 N_f T \frac{P^2}{|{\bf p}|^2}\int_0^\infty \frac{dk}{2\pi} n_F(k T)\left(1-\frac{1}{\sqrt{P^2}}{\rm Re}\sqrt{P^2-4 k^2 T^2-4 i p_0 k T}\right)\nonumber\\
  \Pi_V(P)&=&\frac{e^2 N_f}{8}\sqrt{P^2}\,,
\end{eqnarray}
and where the integration momenta have been scaled by the temperature. Particular care must be taken when evaluating the thermal contributions in the static limit $p_0\rightarrow 0$, finding
\begin{eqnarray}
  \Pi_A-\Pi_V&=&-\frac{e^2 N_f |{\bf p}|}{\pi}\int_0^1 dy n_F(y |{\bf p}|/2)\frac{y^2}{\sqrt{1-y^2}}\,,\nonumber\\
  \Pi_B-\Pi_V&=&\frac{ 2 e^2N_f T \ln 2 }{\pi}+ \frac{e^2 N_f}{\pi} |{\bf p}|\int_0^1 dy n_F(y |{\bf p}|/2)\sqrt{1-y^2}\,.
\end{eqnarray}
One recognizes the 2+1 dimensional Debye mass
\begin{equation}
  m_D^2\equiv \frac{2 e^2 N_f T \ln 2 }{\pi}\,,
\end{equation}
in the zero momentum limit of $\Pi_B-\Pi_V$.

The fermion contribution and the remaining ghost contribution are easy to evaluate:
\begin{eqnarray}
  \label{eq:f1s}
  f_{\rm ghost}&=&-\frac{1}{2}\sumint_{K} \ln K^2=-\int \frac{d^2 k}{(2\pi)^2}\ln\left(1-e^{-k/T}\right)=\frac{\zeta(3)T^3}{2 \pi}\,,\nonumber\\
  f_{\rm fermion}&=&-2 N_f \sumint_{\{K\}} \ln K^2=-4 N_f \int \frac{d^2k}{(2 \pi)^2}\ln\left(1+e^{-k/T}\right)=-\frac{3 N_f \zeta(3) T^3}{2\pi}\,,
\end{eqnarray}
where only the matter contribution of the thermal sums give non-vanishing contributions. For the photons, we note that
\begin{equation}
  \label{eq:partA}
  f_{A,B}=\frac{1}{2}\sumint_K \ln \left(K^2+\Pi_{A,B}(K)\right)=\frac{1}{2}\sumint_K \ln \left(K^2+\Pi_V(K)\right)+\frac{1}{2}\sumint_K \ln \left(1+\frac{\Pi_{A,B}(K)-\Pi_V(K)}{K^2+\Pi_V(K)}\right)\,,
\end{equation}
where for large $K$ the asymptotic form of $\Pi_{A,B}-\Pi_V\propto \frac{1}{K^2}$ means that the second term in (\ref{eq:partA}) is both IR- and UV-safe, and thus can be handled numerically.

The remaining term is given by
\begin{eqnarray}
  \label{eq:mainvac}
  f_V&=&\frac{1}{2}\sumint_K \ln \left(K^2+\Pi_V(K)\right)=\frac{1}{4}\sumint_K \ln K^2+\frac{1}{2}\sumint_K \ln \left(\sqrt{K^2}+\frac{\Pi_V(K)}{\sqrt{K^2}}\right)\,,\nonumber\\
  &=&-\frac{\zeta(3)T^3}{4 \pi}+\int \frac{d^DK}{(2\pi)^D}\left(\frac{1}{2}+n_B(i k_0)\right) \ln\left(\sqrt{K^2}+\frac{\Pi_V(K)}{\sqrt{K^2}}\right)\,,
\end{eqnarray}
where $n_B(x)=\frac{1}{e^{x/T}-1}$. The thermal contribution may be rewritten by deforming the contour to run along the Minkowski axis rather than the Euclidean axis because the integrand only has a branch cut, but no singularities anywhere on the principal Riemann sheet \cite{Moore:2002md}. Taking the limit $\epsilon\rightarrow 0$, this leads to
\begin{equation}
f_{V,1}=-\int \frac{d^2k}{(2\pi)^2}\int_k^\infty \frac{d\omega}{\pi}n_B(\omega){\rm Im} \ln\left(\sqrt{-\omega^2+k^2}+\frac{e^2 N_f}{8}\right)\,.
\end{equation}
The contribution may be further simplified as
\begin{eqnarray}
  \label{eq:fv1}
  f_{V,1}&=&-\int_0^\infty \frac{d\omega}{\pi}n_B(\omega)\int_0^{\omega^2} \frac{d(k^2)}{4\pi} {\rm arctan}\left(\frac{8\sqrt{\omega^2-k^2}}{e^2 N_f}\right)\,,\nonumber\\
  &=&-\int_0^\infty \frac{d\omega}{4\pi^2}n_B(\omega)\left[\left(\frac{e^4 N_f^2}{64}+\omega^2\right){\rm arctan}\frac{8 \omega}{e^2 N_f}-\frac{e^2 N_f \omega}{8}\right]\,,
\end{eqnarray}
which is readily evaluated numerically. Alternatively, one may investigate the weak coupling ($\frac{e^2 N_f}{T}\ll1$) and strong coupling ($\frac{e^2 N_f}{T}\gg 1$) limits, which are given by
\begin{eqnarray}
  \lim_{e^2 N_f/T\rightarrow 0} f_{V,1}&=&-\frac{\zeta(3)T^3}{4\pi}+\frac{e^2 N_f T^2}{96}+{\cal O}(e^4 N_f^2 T)\,,\nonumber\\
  \lim_{e^2 N_f/T\rightarrow \infty} f_{V,1}&=&-\frac{4 \pi^2 T^4}{45 e^2 N_f}+{\cal O}\left(T^5/(e^4 N_f^2)\right)\,.
\end{eqnarray}
From this one recovers what could already have been gleaned from the original sum-integral representation in (\ref{eq:mainvac}): for weak coupling where $\Pi_V\rightarrow 0$, $f_{V}=-\frac{\zeta(3) T^3}{2\pi}$, corresponding to the free energy density of a single bosonic degree of freedom; conversely, for strong coupling where $K^2+\Pi_V\simeq \Pi_V$, the $f_{V,1}$ contribution vanishes and $f_{V}=-\frac{\zeta(3) T^3}{4\pi}$, corresponding to $\frac{1}{2}$ degree of freedom.

\subsection{Apparently divergent vacuum energy at four-loop order}
\label{sec:vacdiv}

Finally, let us discuss the vacuum contribution
\begin{equation}
  \label{eq:fv2first}
f_{V,2}=\frac{1}{2}\int \frac{d^DK}{(2\pi)^D}\ln\left(\sqrt{K^2}+\frac{\Pi_V(K)}{\sqrt{K^2}}\right)\,,
\end{equation}
which vanishes identically for both $e^2 N_f=0,\infty$. However, at face value $f_{V,2}$ includes a logarithmic divergence for any finite value of $e^2 N_f$. This divergence arises at four-loop order in a perturbative expansion, which can be seen by expanding (\ref{eq:fv2first}) in powers of $\Pi_V\propto e^2 N_f \sqrt{K^2}$ such that $f_{V,2}\propto (e^2 N_f)^3 \int \frac{d^D K}{(K^2)^{3/2}}\propto \frac{(e^2 N_f)^3}{\epsilon}$.

The appearance of the $e^6$ coefficient is similar to the $g^6$ infrared divergence encountered for non-abelian gauge theories, also known as the ``Linde problem'' \cite{Linde:1980ts}. However, we believe these issues are unrelated because for the case of QED3, the apparent divergence is in the ultraviolet, not in the infrared. 
The naively UV divergent contribution to $f_{V,2}$ may be calculated by considering
\begin{equation}
  g_{V,2}=\frac{1}{2}\int \frac{d^DK}{(2\pi)^D}\frac{1}{\left(K^2\right)^{1/2+\epsilon}+c }=\frac{c^2}{32 \pi^2}\left[\frac{1}{\epsilon}+2-\gamma_E+\ln \frac{\mu^2 \pi}{c^8}+{\cal O}(\epsilon)\right]\,,
\end{equation}
in dimensional regularization where $c=\mu^{2 \epsilon}\frac{8 e^2 N_f}{(4\pi)^{D/2}}\Gamma(2-\frac{D}{2})\frac{\Gamma^2(D/2)}{\Gamma(D)}$ from (\ref{eqpiv11}) and $\gamma_E$ is Euler's constant. Subsequently integrating w.r.t. $c$ we find
\begin{equation}
  \label{eq:fv2}
f_{V,2}=\frac{1}{96\pi^2}\left(\frac{e^2 N_f}{8}\right)^3\left[\frac{1}{\epsilon}+\frac{5}{3}+4 \ln \frac{\bar\mu^2}{e^4 N_f^2/128} +{\cal O}(\epsilon) \right]\,,
\end{equation}
where $\bar\mu=\sqrt{4 \pi e^{-\gamma_E}}\mu $ is the $\overline{\rm MS}$ scheme renormalization scale. The apparently divergent contribution to the free energy density at four-loop ($e^6$) is problematic: since there are no divergences requiring renormalization for the charge, mass or wave-function, the only way to cancel the divergence would be by adding a vacuum-energy counterterm to the Lagrangian. However, even after doing so, this would imply that the vacuum energy thus found is renormalization-scale dependent, since there are no other divergences to cancel the non-vanishing derivative $\frac{\partial f_{V}}{\partial \bar\mu}$. Since the free energy is a physical observable, this cannot happen.

Further inspection reveals that the problem lies with the naive $N_f\rightarrow \infty$ limit. It is possible to consider further corrections to the photon polarization tensor which are formally suppressed by powers of $N_f$, for instance at the two-loop level, cf. Ref.~\cite{Grozin:2005yg}. One two-loop contribution (which by itself is not gauge invariant) originates from a non-vanishing fermion self-energy, modifying the fermion propagator as
\begin{equation}
  \Delta^{-1}(K)\rightarrow i\slashed{K}\left(1+\Sigma(K)\right)\,.
  \end{equation}
To leading order in large $N_f$, $\Sigma(K)$ may be calculated by using the resummed photon propagator to find
\begin{equation}
\Sigma(K)\propto  \frac{1}{N_f} \ln \frac{e^4 N_f^2}{K^2}\,,
\end{equation}
which in turn suggests that similar contributions of higher order in $N_f$ may be non-perturbatively resummed to give \cite{Pisarski:1984dj}
\begin{equation}
  1+\Sigma(K)=\left(c_0\times \frac{e^2 N_f}{\sqrt{K^2}}\right)^{\frac{8}{N_f \pi^2}}\,,
\end{equation}
with a calculable constant $c_0$. Including the self-energy correction into the evaluation for the photon polarization tensor (\ref{eq:matterpi}) then suggests the modification
\begin{eqnarray}
  \label{eq:lnf}
  \Pi_V(K)=\frac{e^2 N_f}{8}\sqrt{K^2}\rightarrow \left(\frac{e^2 N_f}{8}\right)^{1-\frac{8}{N_f \pi^2}}\left(K^2\right)^{1/2+\frac{4}{N_f\pi^2}}\,,\nonumber\\
  \Pi_{A,B}-\Pi_V\propto \frac{e^2 N_f}{8} T \rightarrow \left(\frac{e^2 N_f}{8}\right)^{1-\frac{8}{N_f \pi^2}} T^{1+\frac{8}{N_f \pi^2}}\,.
  \end{eqnarray}
While these modifications do not modify most of the results for the free energy discussed above at the ${\cal O}(N_f^0)$ level, there are two notable exceptions. 

First, consider the contribution $f_{V,2}$ in light of these non-perturbative resummations of formally sub-leading $1/N_f$ corrections. Expanding (\ref{eq:fv2first}) in powers of $\Pi_V$ as before, but with $\Pi_V\propto \sqrt{K^2}^{1+\frac{8}{N_f \pi^2}}$ one finds that the four-loop perturbative expression is finite in dimensional regularization because $\frac{8}{N_f\pi^2}$ takes over the role of $\epsilon$. Hence we find
\begin{equation}
  f_{V,2}\rightarrow \frac{1}{2}\int\frac{d^D K}{(2\pi)^D}\ln\left[\left(K^2\right)^{1/2-\frac{4}{N_f \pi^2}}+\left(\frac{e^2 N_f}{8}\right)^{1-\frac{8}{N_f \pi^2}}\right]\propto \left(\frac{e^2 N_f}{8}\right)^3 \times N_f\,.
\end{equation}
Thus the result of including the naively sub-leading terms in the $1/N_f$ expansion is that the apparent UV divergence of the free energy gets turned into a finite contribution to order ${\cal O}(N_f)$. Therefore, after resummation, the vacuum free energy is no longer renormalization-scale dependent, but there is a non-vanishing and finite cosmological constant contribution at order ${\cal O}(e^6 N_f^4)$.

\subsection{Suppression of in-medium tensor contributions at strong coupling}
\label{sec:msup}

The second instance where the formally sub-leading corrections (\ref{eq:lnf}) become important is in the numerical evaluation of the in-medium contribution in Eq.~(\ref{eq:partA}) near zero temperature. Specifically, without taking (\ref{eq:lnf}) into account, the temperature-dependence for the polarization tensor components (\ref{piexp}) may be scaled out by taking $P\rightarrow T \hat{P}$, $e^2\rightarrow T \hat{e^2}$. As a consequence, one would expect the in-medium contributions to $\Pi_{A,B}$ to have non-vanishing contributions to the free energy $f_{A,B}$ even in the zero temperature limit.

However, taking into account (\ref{eq:lnf}), our calculation with naive in-medium contributions can only be trusted in a regime where
\begin{equation}
  \frac{e^2 N_f}{T}\ll e^{N_f \pi^2/8}\,,
\end{equation}
whereas for $\frac{e^2 N_f}{T}\rightarrow \infty$, the ${\cal O}(N_f^0)$ photon contribution to the free energy is given by $f_V$ in (\ref{eq:mainvac}).

\subsection{Numerical evaluation of thermal contribution}

The thermal photon polarization tensor contribution to the free energy is handled fully numerically by directly evaluating
\begin{equation}
  f_{A,B}^{(M)}=\frac{1}{2}\sumint_K \ln \left(1+\frac{\Pi_{A,B}(K)-\Pi_V(K)}{K^2+\Pi_V(K)}\right)\,.
\end{equation}
Specifically, this is done by performing the sum over Matsubara frequencies and using Gauss-Legendre quadrature for the remaining integral as in Refs.~\cite{Romatschke:2019rjk,Romatschke:2019wxc}:
\begin{equation}
  \label{eq:fabm}
  f_{A,B}^{(M)}=\frac{T^3}{16}\sum_{n=-M}^M\sum_{i=1}^N W_i \ln \left(1+\frac{\Pi_{A,B}(\omega_n,k_i T)-\Pi_V(\omega_n,k_i T)}{\omega_n^2+\Pi_V(\omega_n,k_i T)}\right)\,,
\end{equation}
where we used $k_i=\sqrt{\tan\left(\frac{x_i \pi}{2}\right)}+0^+$ to compactify the infinite interval including a small regulator to avoid any IR divergences. Here $x_i,W_i$ are the nodes and modified weights, respectively, defined by the roots of the Legendre polynomial of order N:
\begin{equation}
  P_N(x_i)=0\,,\quad W_i=\frac{1}{(1-x_i^2)\left(P_N^\prime(x_i)\right) \cos^2\left(\frac{x_i \pi}{2}\right)}\,.
  \end{equation}
In practice, because of the symmetries of the integrand, only nodes with $n\geq 0$, $x_i\geq 0$ need to be summed over.
Tabulated values for $x_i$ can be easily generated with high precision for $N$ up to $N\simeq 2000$, but in practice $N\simeq 200$ seems sufficient to obtain percent level precision.

We note that, in practice, we find that $f_A\leq 0$ and $f_B\geq 0$ and of similar magnitude for all values of the coupling. The numerical code for obtaining $f_{A,B}^{(M)}$ and $f_{V,1}$ as well as tabulated numerical results are publicly available at \cite{codedown}.

\section{Results and Discussion}


The full free energy density for QED3 in the large $N_f$ limit is given by
\begin{equation}
  f_{\rm QED3}=f_{\rm fermion}+f_{\rm ghost}+f_{\rm photon}\,,
  \end{equation}
    where $f_{\rm fermion}$ and $f_{\rm ghost}$ are given in Eq.~(\ref{eq:f1s}). Here $f_{\rm photon}=f_A+f_B$ with
    \begin{equation}
      f_{A,B}=-\frac{\zeta(3) T^3}{4\pi}+f_{V,1}+f_{V,2}+f_{A,B}^{(M)}
    \end{equation}
    where $f_{V,1},f_{V,2}$ and the matter contributions $f_{A,B}^{(M)}$ are given in Eqns.~(\ref{eq:fv1}), (\ref{eq:fv2}) and (\ref{eq:fabm}), respectively. As pointed out in section \ref{sec:vacdiv}, $f_{V,2}$ is UV-divergent in the naive large $N_f$ limit, with the expectation that this divergence gets turned into a finite ${\cal O}(N_f)$ contribution once higher order terms in $1/N_f$ are resummed. Since this resummation is beyond the scope of the present work, we focus on the difference between vacuum and finite-temperature quantities where $f_{V,2}$ drops out. In particular, we study the pressure (minus the free energy density) difference
    \begin{equation}
      \label{eq:npres}
      \frac{P(T)-P(0)}{\frac{\zeta(3)T^3}{2 \pi}}=3 N_f-2 f_{V,1}-f_{A}^{(M)}-f_{B}^{(M)}\,,
    \end{equation}
    where we have normalized the pressure to the pressure of a free (non-interacting) bosonic degree of freedom. Note that the normalized pressure is $3 N_f$ and not $4 N_f$ because in three dimensions each fermionic degree of freedom contributes only $\frac{3}{4}$ of a bosonic degree of freedom.

\begin{figure*}[t]
  \includegraphics[width=0.8\linewidth]{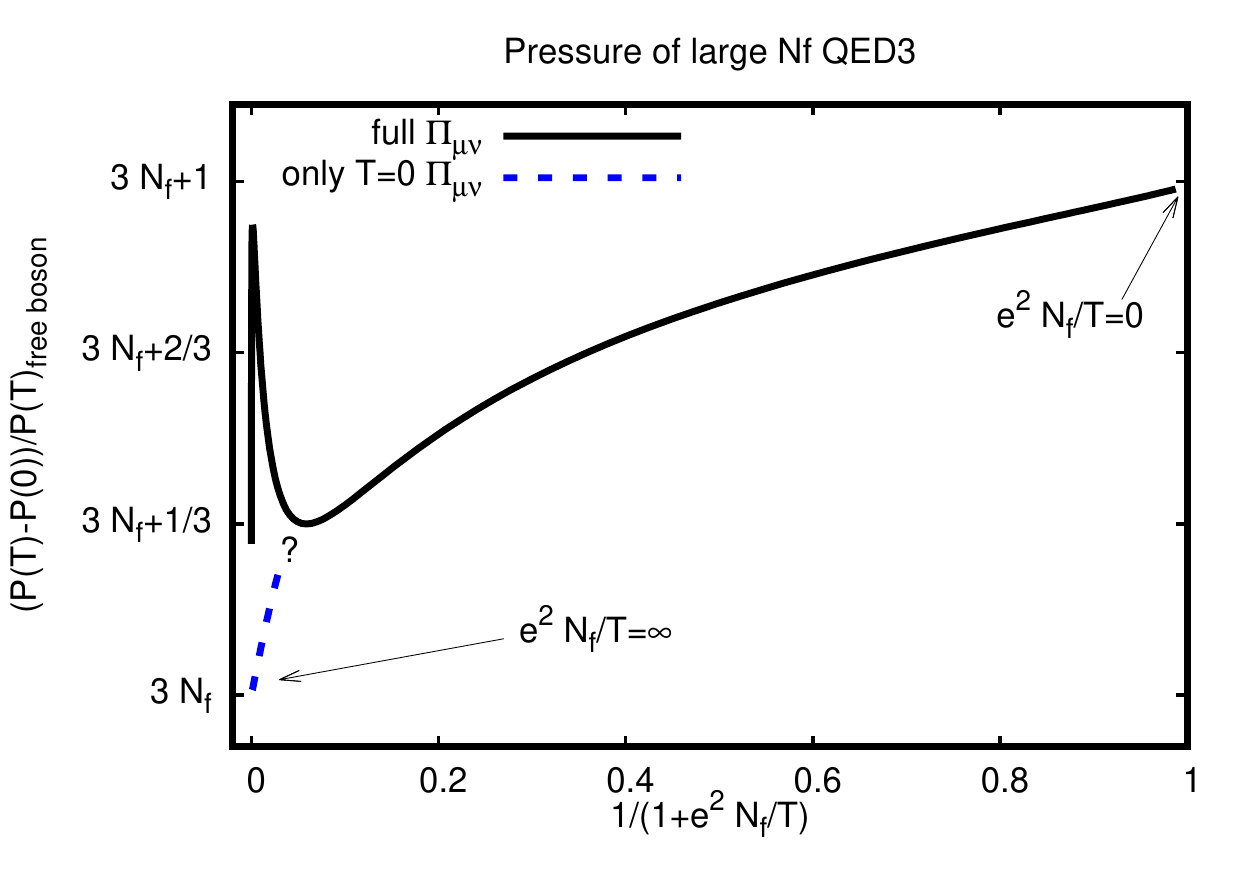}
  \caption{\label{fig1} Normalized pressure (\ref{eq:npres}) in large $N_f$ QED in 2+1d for all coupling values (full line). For comparison, the normalized pressure using only the vacuum polarization tensor is also shown (dashed line). Horizontal axis has been compactified in order to show the full range $\frac{e^2 N_f}{T} \in [0,\infty)$. Arrows indicate weak-coupling and infinite coupling limit, respectively. The question mark indicates that we do not trust our results using the full in-medium polarization tensor to be a good approximation at large, but fixed $N_f$ in this region. See text for details.}
\end{figure*}
    
    As discussed above, results for $f_{V,1},f_{A}^{(M)},f_{B}^{(M)}$ can be obtained numerically for arbitrary values of $\frac{e^2 N_f}{T}$. However, for reasons discussed in section \ref{sec:msup}, for any finite $N_f$ we expect contributions that are naively higher order in $1/N_f$ to suppress the in-medium contributions to $\Pi_{A,B}$ for sufficiently low temperatures/high values of $\frac{e^2 N_f}{T}$. Therefore, we expect our numerically obtained results for $f_{A,B}^{(M)}$ to lose validity at a large but finite value of $\frac{e^2 N_f}{T}$. Following the arguments in Ref.~\cite{Romatschke:2019mjm}, we expect the  $\frac{e^2 N_f}{T}\rightarrow \infty$ limit of the pressure to be well approximated by neglecting $f_{A,B}^{(M)}$, but including $f_{V,1}$.

    Our main result for the pressure is shown in Fig.~\ref{fig1}. For weak coupling values $\frac{e^2 N_f}{T}\ll 1$, we find that the normalized pressure decreases monotonically from the free theory value at $3 N_f+1$. This trend continues up to coupling values of approximately $\frac{e^2 N_f}{T}\leq 16$, at which point the normalized pressure (\ref{eq:npres}) is numerically given by $3N_f+0.333(3)$. For $\frac{e^2 N_f}{T}\geq 16$, the normalized pressure then starts to rise as a function of coupling, similar to what has been reported in the case of QED4 in Refs.~\cite{Moore:2002md,Ipp:2003zr} (see Fig.~\ref{fig1}). (Note that apparent non-monotonic behavior shown in Fig.~\ref{fig1} is a result of the normalization used for plotting; the (un-normalized) pressure itself is always monotonically increasing with temperature as it should.) Eventually, the normalized pressure hits a maximum below $3 N_f+1$ and starts to decrease again for $\frac{e^2 N_f}{T}\geq 100$, with the numerical evaluation of $f_{A,B}^{(M)}$ becoming more challenging in this region. 

    Based on the arguments given in section \ref{sec:msup}, we suspect that for fixed, but large $N_f$, the normalized pressure for $\frac{e^2 N_f}{T}\gg 16$ may continue to decrease towards $3 N_f$, departing from our calculation that is using the in-medium polarization tensor evaluated in the naive large $N_f$ limit. This is indicated by a question mark and the result using only the vacuum polarization tensor shown in Fig.~\ref{fig1}. We would invite follow-up studies from lattice gauge theory simulations at finite temperature in particular for $\frac{e^2 N_f}{T}\geq 16$ to settle this issue.

  \section{Acknowledgments}

  This work was supported by the Department of Energy, DOE award No DE-SC0017905. We would like to thank T.~DeGrand, A.~Hasenfratz, M.~Laine, R.~Pisarski, Y.~Schr\"oder, B.~Svetitsky and A.~Vuorinen for helpful discussions.

\bibliography{cft}

\begin{thebibliography}{35}%
\makeatletter
\providecommand \@ifxundefined [1]{%
 \@ifx{#1\undefined}
}%
\providecommand \@ifnum [1]{%
 \ifnum #1\expandafter \@firstoftwo
 \else \expandafter \@secondoftwo
 \fi
}%
\providecommand \@ifx [1]{%
 \ifx #1\expandafter \@firstoftwo
 \else \expandafter \@secondoftwo
 \fi
}%
\providecommand \natexlab [1]{#1}%
\providecommand \enquote  [1]{``#1''}%
\providecommand \bibnamefont  [1]{#1}%
\providecommand \bibfnamefont [1]{#1}%
\providecommand \citenamefont [1]{#1}%
\providecommand \href@noop [0]{\@secondoftwo}%
\providecommand \href [0]{\begingroup \@sanitize@url \@href}%
\providecommand \@href[1]{\@@startlink{#1}\@@href}%
\providecommand \@@href[1]{\endgroup#1\@@endlink}%
\providecommand \@sanitize@url [0]{\catcode `\\12\catcode `\$12\catcode
  `\&12\catcode `\#12\catcode `\^12\catcode `\_12\catcode `\%12\relax}%
\providecommand \@@startlink[1]{}%
\providecommand \@@endlink[0]{}%
\providecommand \url  [0]{\begingroup\@sanitize@url \@url }%
\providecommand \@url [1]{\endgroup\@href {#1}{\urlprefix }}%
\providecommand \urlprefix  [0]{URL }%
\providecommand \Eprint [0]{\href }%
\providecommand \doibase [0]{http://dx.doi.org/}%
\providecommand \selectlanguage [0]{\@gobble}%
\providecommand \bibinfo  [0]{\@secondoftwo}%
\providecommand \bibfield  [0]{\@secondoftwo}%
\providecommand \translation [1]{[#1]}%
\providecommand \BibitemOpen [0]{}%
\providecommand \bibitemStop [0]{}%
\providecommand \bibitemNoStop [0]{.\EOS\space}%
\providecommand \EOS [0]{\spacefactor3000\relax}%
\providecommand \BibitemShut  [1]{\csname bibitem#1\endcsname}%
\let\auto@bib@innerbib\@empty
\bibitem [{\citenamefont {Maldacena}(1999)}]{Maldacena:1997re}%
  \BibitemOpen
  \bibfield  {author} {\bibinfo {author} {\bibfnamefont {Juan~Martin}\
  \bibnamefont {Maldacena}},\ }\bibfield  {title} {\enquote {\bibinfo {title}
  {{The Large N limit of superconformal field theories and supergravity}},}\
  }\href {\doibase 10.1023/A:1026654312961, 10.4310/ATMP.1998.v2.n2.a1}
  {\bibfield  {journal} {\bibinfo  {journal} {Int. J. Theor. Phys.}\ }\textbf
  {\bibinfo {volume} {38}},\ \bibinfo {pages} {1113--1133} (\bibinfo {year}
  {1999})},\ \bibinfo {note} {[Adv. Theor. Math. Phys.2,231(1998)]},\ \Eprint
  {http://arxiv.org/abs/hep-th/9711200} {arXiv:hep-th/9711200} \BibitemShut
  {NoStop}%
\bibitem [{\citenamefont {Gubser}\ \emph {et~al.}(1998)\citenamefont {Gubser},
  \citenamefont {Klebanov},\ and\ \citenamefont {Tseytlin}}]{Gubser:1998nz}%
  \BibitemOpen
  \bibfield  {author} {\bibinfo {author} {\bibfnamefont {Steven~S.}\
  \bibnamefont {Gubser}}, \bibinfo {author} {\bibfnamefont {Igor~R.}\
  \bibnamefont {Klebanov}}, \ and\ \bibinfo {author} {\bibfnamefont
  {Arkady~A.}\ \bibnamefont {Tseytlin}},\ }\bibfield  {title} {\enquote
  {\bibinfo {title} {{Coupling constant dependence in the thermodynamics of N=4
  supersymmetric Yang-Mills theory}},}\ }\href {\doibase
  10.1016/S0550-3213(98)00514-8} {\bibfield  {journal} {\bibinfo  {journal}
  {Nucl. Phys.}\ }\textbf {\bibinfo {volume} {B534}},\ \bibinfo {pages}
  {202--222} (\bibinfo {year} {1998})},\ \Eprint
  {http://arxiv.org/abs/hep-th/9805156} {arXiv:hep-th/9805156} \BibitemShut
  {NoStop}%
\bibitem [{\citenamefont {Itzhaki}\ \emph {et~al.}(1998)\citenamefont
  {Itzhaki}, \citenamefont {Maldacena}, \citenamefont {Sonnenschein},\ and\
  \citenamefont {Yankielowicz}}]{Itzhaki:1998dd}%
  \BibitemOpen
  \bibfield  {author} {\bibinfo {author} {\bibfnamefont {Nissan}\ \bibnamefont
  {Itzhaki}}, \bibinfo {author} {\bibfnamefont {Juan~Martin}\ \bibnamefont
  {Maldacena}}, \bibinfo {author} {\bibfnamefont {Jacob}\ \bibnamefont
  {Sonnenschein}}, \ and\ \bibinfo {author} {\bibfnamefont {Shimon}\
  \bibnamefont {Yankielowicz}},\ }\bibfield  {title} {\enquote {\bibinfo
  {title} {{Supergravity and the large N limit of theories with sixteen
  supercharges}},}\ }\href {\doibase 10.1103/PhysRevD.58.046004} {\bibfield
  {journal} {\bibinfo  {journal} {Phys. Rev.}\ }\textbf {\bibinfo {volume}
  {D58}},\ \bibinfo {pages} {046004} (\bibinfo {year} {1998})},\ \Eprint
  {http://arxiv.org/abs/hep-th/9802042} {arXiv:hep-th/9802042} \BibitemShut
  {NoStop}%
\bibitem [{\citenamefont {Policastro}\ \emph {et~al.}(2001)\citenamefont
  {Policastro}, \citenamefont {Son},\ and\ \citenamefont
  {Starinets}}]{Policastro:2001yc}%
  \BibitemOpen
  \bibfield  {author} {\bibinfo {author} {\bibfnamefont {G.}~\bibnamefont
  {Policastro}}, \bibinfo {author} {\bibfnamefont {Dan~T.}\ \bibnamefont
  {Son}}, \ and\ \bibinfo {author} {\bibfnamefont {Andrei~O.}\ \bibnamefont
  {Starinets}},\ }\bibfield  {title} {\enquote {\bibinfo {title} {{The Shear
  viscosity of strongly coupled N=4 supersymmetric Yang-Mills plasma}},}\
  }\href {\doibase 10.1103/PhysRevLett.87.081601} {\bibfield  {journal}
  {\bibinfo  {journal} {Phys. Rev. Lett.}\ }\textbf {\bibinfo {volume} {87}},\
  \bibinfo {pages} {081601} (\bibinfo {year} {2001})},\ \Eprint
  {http://arxiv.org/abs/hep-th/0104066} {arXiv:hep-th/0104066 [hep-th]}
  \BibitemShut {NoStop}%
\bibitem [{\citenamefont {Drummond}\ \emph {et~al.}(1998)\citenamefont
  {Drummond}, \citenamefont {Horgan}, \citenamefont {Landshoff},\ and\
  \citenamefont {Rebhan}}]{Drummond:1997cw}%
  \BibitemOpen
  \bibfield  {author} {\bibinfo {author} {\bibfnamefont {I.~T.}\ \bibnamefont
  {Drummond}}, \bibinfo {author} {\bibfnamefont {R.~R.}\ \bibnamefont
  {Horgan}}, \bibinfo {author} {\bibfnamefont {P.~V.}\ \bibnamefont
  {Landshoff}}, \ and\ \bibinfo {author} {\bibfnamefont {A.}~\bibnamefont
  {Rebhan}},\ }\bibfield  {title} {\enquote {\bibinfo {title} {{Foam diagram
  summation at finite temperature}},}\ }\href {\doibase
  10.1016/S0550-3213(98)00210-7} {\bibfield  {journal} {\bibinfo  {journal}
  {Nucl. Phys.}\ }\textbf {\bibinfo {volume} {B524}},\ \bibinfo {pages}
  {579--600} (\bibinfo {year} {1998})},\ \Eprint
  {http://arxiv.org/abs/hep-ph/9708426} {arXiv:hep-ph/9708426} \BibitemShut
  {NoStop}%
\bibitem [{\citenamefont
  {Romatschke}(2019{\natexlab{a}})}]{Romatschke:2019ybu}%
  \BibitemOpen
  \bibfield  {author} {\bibinfo {author} {\bibfnamefont {Paul}\ \bibnamefont
  {Romatschke}},\ }\bibfield  {title} {\enquote {\bibinfo {title} {{Finite
  temperature CFT results for all couplings: O(N) model in 2+1 dimensions}},}\
  }\href@noop {} {\  (\bibinfo {year} {2019}{\natexlab{a}})},\ \Eprint
  {http://arxiv.org/abs/1904.09995} {arXiv:1904.09995} \BibitemShut {NoStop}%
\bibitem [{\citenamefont {DeWolfe}\ and\ \citenamefont
  {Romatschke}(2019)}]{DeWolfe:2019etx}%
  \BibitemOpen
  \bibfield  {author} {\bibinfo {author} {\bibfnamefont {Oliver}\ \bibnamefont
  {DeWolfe}}\ and\ \bibinfo {author} {\bibfnamefont {Paul}\ \bibnamefont
  {Romatschke}},\ }\bibfield  {title} {\enquote {\bibinfo {title} {{Strong
  Coupling Universality at Large N for Pure CFT Thermodynamics in 2+1
  dimensions}},}\ }\href@noop {} {\  (\bibinfo {year} {2019})},\ \Eprint
  {http://arxiv.org/abs/1905.06355} {arXiv:1905.06355 [hep-th]} \BibitemShut
  {NoStop}%
\bibitem [{\citenamefont
  {Romatschke}(2019{\natexlab{b}})}]{Romatschke:2019gck}%
  \BibitemOpen
  \bibfield  {author} {\bibinfo {author} {\bibfnamefont {Paul}\ \bibnamefont
  {Romatschke}},\ }\bibfield  {title} {\enquote {\bibinfo {title} {{Analytic
  Transport from Weak to Strong Coupling in the O(N) model}},}\ }\href@noop {}
  {\  (\bibinfo {year} {2019}{\natexlab{b}})},\ \Eprint
  {http://arxiv.org/abs/1905.09290} {arXiv:1905.09290 [hep-th]} \BibitemShut
  {NoStop}%
\bibitem [{\citenamefont {Moore}(2002)}]{Moore:2002md}%
  \BibitemOpen
  \bibfield  {author} {\bibinfo {author} {\bibfnamefont {Guy~D.}\ \bibnamefont
  {Moore}},\ }\bibfield  {title} {\enquote {\bibinfo {title} {{Pressure of hot
  QCD at large N(f)}},}\ }\href {\doibase 10.1088/1126-6708/2002/10/055}
  {\bibfield  {journal} {\bibinfo  {journal} {JHEP}\ }\textbf {\bibinfo
  {volume} {10}},\ \bibinfo {pages} {055} (\bibinfo {year} {2002})},\ \Eprint
  {http://arxiv.org/abs/hep-ph/0209190} {arXiv:hep-ph/0209190 [hep-ph]}
  \BibitemShut {NoStop}%
\bibitem [{\citenamefont {Ipp}\ \emph {et~al.}(2003)\citenamefont {Ipp},
  \citenamefont {Moore},\ and\ \citenamefont {Rebhan}}]{Ipp:2003zr}%
  \BibitemOpen
  \bibfield  {author} {\bibinfo {author} {\bibfnamefont {Andreas}\ \bibnamefont
  {Ipp}}, \bibinfo {author} {\bibfnamefont {Guy~D.}\ \bibnamefont {Moore}}, \
  and\ \bibinfo {author} {\bibfnamefont {Anton}\ \bibnamefont {Rebhan}},\
  }\bibfield  {title} {\enquote {\bibinfo {title} {{Comment on and erratum to
  `Pressure of hot QCD at large N(f)'}},}\ }\href {\doibase
  10.1088/1126-6708/2003/01/037} {\bibfield  {journal} {\bibinfo  {journal}
  {JHEP}\ }\textbf {\bibinfo {volume} {01}},\ \bibinfo {pages} {037} (\bibinfo
  {year} {2003})},\ \Eprint {http://arxiv.org/abs/hep-ph/0301057}
  {arXiv:hep-ph/0301057 [hep-ph]} \BibitemShut {NoStop}%
\bibitem [{\citenamefont {D'Hoker}(1982)}]{DHoker:1981bjo}%
  \BibitemOpen
  \bibfield  {author} {\bibinfo {author} {\bibfnamefont {Eric}\ \bibnamefont
  {D'Hoker}},\ }\bibfield  {title} {\enquote {\bibinfo {title} {{PERTURBATIVE
  RESULTS ON QCD in three-dimensions AT FINITE TEMPERATURE}},}\ }\href
  {\doibase 10.1016/0550-3213(82)90441-2} {\bibfield  {journal} {\bibinfo
  {journal} {Nucl. Phys.}\ }\textbf {\bibinfo {volume} {B201}},\ \bibinfo
  {pages} {401--428} (\bibinfo {year} {1982})}\BibitemShut {NoStop}%
\bibitem [{\citenamefont {Pisarski}(1984)}]{Pisarski:1984dj}%
  \BibitemOpen
  \bibfield  {author} {\bibinfo {author} {\bibfnamefont {Robert~D.}\
  \bibnamefont {Pisarski}},\ }\bibfield  {title} {\enquote {\bibinfo {title}
  {{Chiral Symmetry Breaking in Three-Dimensional Electrodynamics}},}\ }\href
  {\doibase 10.1103/PhysRevD.29.2423} {\bibfield  {journal} {\bibinfo
  {journal} {Phys. Rev.}\ }\textbf {\bibinfo {volume} {D29}},\ \bibinfo {pages}
  {2423} (\bibinfo {year} {1984})}\BibitemShut {NoStop}%
\bibitem [{\citenamefont {Appelquist}\ \emph {et~al.}(1988)\citenamefont
  {Appelquist}, \citenamefont {Nash},\ and\ \citenamefont
  {Wijewardhana}}]{Appelquist:1988sr}%
  \BibitemOpen
  \bibfield  {author} {\bibinfo {author} {\bibfnamefont {Thomas}\ \bibnamefont
  {Appelquist}}, \bibinfo {author} {\bibfnamefont {Daniel}\ \bibnamefont
  {Nash}}, \ and\ \bibinfo {author} {\bibfnamefont {L.~C.~R.}\ \bibnamefont
  {Wijewardhana}},\ }\bibfield  {title} {\enquote {\bibinfo {title} {{Critical
  Behavior in (2+1)-Dimensional QED}},}\ }\href {\doibase
  10.1103/PhysRevLett.60.2575} {\bibfield  {journal} {\bibinfo  {journal}
  {Phys. Rev. Lett.}\ }\textbf {\bibinfo {volume} {60}},\ \bibinfo {pages}
  {2575} (\bibinfo {year} {1988})}\BibitemShut {NoStop}%
\bibitem [{\citenamefont {Azcoiti}\ and\ \citenamefont
  {Luo}(1993)}]{Azcoiti:1993ey}%
  \BibitemOpen
  \bibfield  {author} {\bibinfo {author} {\bibfnamefont {Vicente}\ \bibnamefont
  {Azcoiti}}\ and\ \bibinfo {author} {\bibfnamefont {Xiang-Qian}\ \bibnamefont
  {Luo}},\ }\bibfield  {title} {\enquote {\bibinfo {title} {{Phase structure of
  compact lattice QED in three-dimensions with massless Fermions}},}\ }\href
  {\doibase 10.1142/S0217732393002373} {\bibfield  {journal} {\bibinfo
  {journal} {Mod. Phys. Lett.}\ }\textbf {\bibinfo {volume} {A8}},\ \bibinfo
  {pages} {3635--3642} (\bibinfo {year} {1993})},\ \Eprint
  {http://arxiv.org/abs/hep-lat/9212011} {arXiv:hep-lat/9212011 [hep-lat]}
  \BibitemShut {NoStop}%
\bibitem [{\citenamefont {Lee}\ and\ \citenamefont {Maris}(2003)}]{Lee:2002id}%
  \BibitemOpen
  \bibfield  {author} {\bibinfo {author} {\bibfnamefont {Dean}\ \bibnamefont
  {Lee}}\ and\ \bibinfo {author} {\bibfnamefont {Pieter}\ \bibnamefont
  {Maris}},\ }\bibfield  {title} {\enquote {\bibinfo {title} {{Massless QED(3)
  with explicit fermions}},}\ }\href {\doibase 10.1103/PhysRevD.67.076002}
  {\bibfield  {journal} {\bibinfo  {journal} {Phys. Rev.}\ }\textbf {\bibinfo
  {volume} {D67}},\ \bibinfo {pages} {076002} (\bibinfo {year} {2003})},\
  \Eprint {http://arxiv.org/abs/hep-lat/0212033} {arXiv:hep-lat/0212033
  [hep-lat]} \BibitemShut {NoStop}%
\bibitem [{\citenamefont {Hands}\ \emph {et~al.}(2002)\citenamefont {Hands},
  \citenamefont {Kogut},\ and\ \citenamefont {Strouthos}}]{Hands:2002dv}%
  \BibitemOpen
  \bibfield  {author} {\bibinfo {author} {\bibfnamefont {S.~J.}\ \bibnamefont
  {Hands}}, \bibinfo {author} {\bibfnamefont {J.~B.}\ \bibnamefont {Kogut}}, \
  and\ \bibinfo {author} {\bibfnamefont {C.~G.}\ \bibnamefont {Strouthos}},\
  }\bibfield  {title} {\enquote {\bibinfo {title} {{Noncompact QED(3) with N(f)
  greater than or equal to 2}},}\ }\href {\doibase
  10.1016/S0550-3213(02)00869-6} {\bibfield  {journal} {\bibinfo  {journal}
  {Nucl. Phys.}\ }\textbf {\bibinfo {volume} {B645}},\ \bibinfo {pages}
  {321--336} (\bibinfo {year} {2002})},\ \Eprint
  {http://arxiv.org/abs/hep-lat/0208030} {arXiv:hep-lat/0208030 [hep-lat]}
  \BibitemShut {NoStop}%
\bibitem [{\citenamefont {Strouthos}\ and\ \citenamefont
  {Kogut}(2007)}]{Strouthos:2008kc}%
  \BibitemOpen
  \bibfield  {author} {\bibinfo {author} {\bibfnamefont {Costas}\ \bibnamefont
  {Strouthos}}\ and\ \bibinfo {author} {\bibfnamefont {John~B.}\ \bibnamefont
  {Kogut}},\ }\bibfield  {title} {\enquote {\bibinfo {title} {{The Phases of
  Non-Compact QED(3)}},}\ }\bibfield  {booktitle} {\emph {\bibinfo {booktitle}
  {{Proceedings, 25th International Symposium on Lattice field theory (Lattice
  2007): Regensburg, Germany, July 30-August 4, 2007}}},\ }\href {\doibase
  10.22323/1.042.0278} {\bibfield  {journal} {\bibinfo  {journal} {PoS}\
  }\textbf {\bibinfo {volume} {LATTICE2007}},\ \bibinfo {pages} {278} (\bibinfo
  {year} {2007})},\ \Eprint {http://arxiv.org/abs/0804.0300} {arXiv:0804.0300
  [hep-lat]} \BibitemShut {NoStop}%
\bibitem [{\citenamefont {Raviv}\ \emph {et~al.}(2014)\citenamefont {Raviv},
  \citenamefont {Shamir},\ and\ \citenamefont {Svetitsky}}]{Raviv:2014xna}%
  \BibitemOpen
  \bibfield  {author} {\bibinfo {author} {\bibfnamefont {Ohad}\ \bibnamefont
  {Raviv}}, \bibinfo {author} {\bibfnamefont {Yigal}\ \bibnamefont {Shamir}}, \
  and\ \bibinfo {author} {\bibfnamefont {Benjamin}\ \bibnamefont {Svetitsky}},\
  }\bibfield  {title} {\enquote {\bibinfo {title} {{Nonperturbative beta
  function in three-dimensional electrodynamics}},}\ }\href {\doibase
  10.1103/PhysRevD.90.014512} {\bibfield  {journal} {\bibinfo  {journal} {Phys.
  Rev.}\ }\textbf {\bibinfo {volume} {D90}},\ \bibinfo {pages} {014512}
  (\bibinfo {year} {2014})},\ \Eprint {http://arxiv.org/abs/1405.6916}
  {arXiv:1405.6916 [hep-lat]} \BibitemShut {NoStop}%
\bibitem [{\citenamefont {Karthik}\ and\ \citenamefont
  {Narayanan}(2016)}]{Karthik:2015sgq}%
  \BibitemOpen
  \bibfield  {author} {\bibinfo {author} {\bibfnamefont {Nikhil}\ \bibnamefont
  {Karthik}}\ and\ \bibinfo {author} {\bibfnamefont {Rajamani}\ \bibnamefont
  {Narayanan}},\ }\bibfield  {title} {\enquote {\bibinfo {title} {{No evidence
  for bilinear condensate in parity-invariant three-dimensional QED with
  massless fermions}},}\ }\href {\doibase 10.1103/PhysRevD.93.045020}
  {\bibfield  {journal} {\bibinfo  {journal} {Phys. Rev.}\ }\textbf {\bibinfo
  {volume} {D93}},\ \bibinfo {pages} {045020} (\bibinfo {year} {2016})},\
  \Eprint {http://arxiv.org/abs/1512.02993} {arXiv:1512.02993 [hep-lat]}
  \BibitemShut {NoStop}%
\bibitem [{\citenamefont {Son}(2015)}]{Son:2015xqa}%
  \BibitemOpen
  \bibfield  {author} {\bibinfo {author} {\bibfnamefont {Dam~Thanh}\
  \bibnamefont {Son}},\ }\bibfield  {title} {\enquote {\bibinfo {title} {{Is
  the Composite Fermion a Dirac Particle?}}}\ }\href {\doibase
  10.1103/PhysRevX.5.031027} {\bibfield  {journal} {\bibinfo  {journal} {Phys.
  Rev.}\ }\textbf {\bibinfo {volume} {X5}},\ \bibinfo {pages} {031027}
  (\bibinfo {year} {2015})},\ \Eprint {http://arxiv.org/abs/1502.03446}
  {arXiv:1502.03446 [cond-mat.mes-hall]} \BibitemShut {NoStop}%
\bibitem [{\citenamefont {Karch}\ and\ \citenamefont
  {Tong}(2016)}]{Karch:2016sxi}%
  \BibitemOpen
  \bibfield  {author} {\bibinfo {author} {\bibfnamefont {Andreas}\ \bibnamefont
  {Karch}}\ and\ \bibinfo {author} {\bibfnamefont {David}\ \bibnamefont
  {Tong}},\ }\bibfield  {title} {\enquote {\bibinfo {title} {{Particle-Vortex
  Duality from 3d Bosonization}},}\ }\href {\doibase 10.1103/PhysRevX.6.031043}
  {\bibfield  {journal} {\bibinfo  {journal} {Phys. Rev.}\ }\textbf {\bibinfo
  {volume} {X6}},\ \bibinfo {pages} {031043} (\bibinfo {year} {2016})},\
  \Eprint {http://arxiv.org/abs/1606.01893} {arXiv:1606.01893 [hep-th]}
  \BibitemShut {NoStop}%
\bibitem [{\citenamefont {Karch}\ \emph {et~al.}(2017)\citenamefont {Karch},
  \citenamefont {Robinson},\ and\ \citenamefont {Tong}}]{Karch:2016aux}%
  \BibitemOpen
  \bibfield  {author} {\bibinfo {author} {\bibfnamefont {Andreas}\ \bibnamefont
  {Karch}}, \bibinfo {author} {\bibfnamefont {Brandon}\ \bibnamefont
  {Robinson}}, \ and\ \bibinfo {author} {\bibfnamefont {David}\ \bibnamefont
  {Tong}},\ }\bibfield  {title} {\enquote {\bibinfo {title} {{More Abelian
  Dualities in 2+1 Dimensions}},}\ }\href {\doibase 10.1007/JHEP01(2017)017}
  {\bibfield  {journal} {\bibinfo  {journal} {JHEP}\ }\textbf {\bibinfo
  {volume} {01}},\ \bibinfo {pages} {017} (\bibinfo {year} {2017})},\ \Eprint
  {http://arxiv.org/abs/1609.04012} {arXiv:1609.04012 [hep-th]} \BibitemShut
  {NoStop}%
\bibitem [{\citenamefont {Kaul}\ and\ \citenamefont
  {Sachdev}(2008)}]{Kaul:2008xw}%
  \BibitemOpen
  \bibfield  {author} {\bibinfo {author} {\bibfnamefont {Ribhu~K.}\
  \bibnamefont {Kaul}}\ and\ \bibinfo {author} {\bibfnamefont {Subir}\
  \bibnamefont {Sachdev}},\ }\bibfield  {title} {\enquote {\bibinfo {title}
  {{Quantum criticality of U(1) gauge theories with fermionic and bosonic
  matter in two spatial dimensions}},}\ }\href {\doibase
  10.1103/PhysRevB.77.155105} {\bibfield  {journal} {\bibinfo  {journal} {Phys.
  Rev.}\ }\textbf {\bibinfo {volume} {B77}},\ \bibinfo {pages} {155105}
  (\bibinfo {year} {2008})},\ \Eprint {http://arxiv.org/abs/0801.0723}
  {arXiv:0801.0723 [cond-mat.str-el]} \BibitemShut {NoStop}%
\bibitem [{\citenamefont {Giombi}\ \emph {et~al.}(2016)\citenamefont {Giombi},
  \citenamefont {Tarnopolsky},\ and\ \citenamefont
  {Klebanov}}]{Giombi:2016fct}%
  \BibitemOpen
  \bibfield  {author} {\bibinfo {author} {\bibfnamefont {Simone}\ \bibnamefont
  {Giombi}}, \bibinfo {author} {\bibfnamefont {Grigory}\ \bibnamefont
  {Tarnopolsky}}, \ and\ \bibinfo {author} {\bibfnamefont {Igor~R.}\
  \bibnamefont {Klebanov}},\ }\bibfield  {title} {\enquote {\bibinfo {title}
  {{On $C_{J}$ and $C_{T}$ in Conformal QED}},}\ }\href {\doibase
  10.1007/JHEP08(2016)156} {\bibfield  {journal} {\bibinfo  {journal} {JHEP}\
  }\textbf {\bibinfo {volume} {08}},\ \bibinfo {pages} {156} (\bibinfo {year}
  {2016})},\ \Eprint {http://arxiv.org/abs/1602.01076} {arXiv:1602.01076
  [hep-th]} \BibitemShut {NoStop}%
\bibitem [{\citenamefont {Franz}\ \emph {et~al.}(2002)\citenamefont {Franz},
  \citenamefont {Tesanovic},\ and\ \citenamefont {Vafek}}]{Franz:2002qy}%
  \BibitemOpen
  \bibfield  {author} {\bibinfo {author} {\bibfnamefont {M.}~\bibnamefont
  {Franz}}, \bibinfo {author} {\bibfnamefont {Z}~\bibnamefont {Tesanovic}}, \
  and\ \bibinfo {author} {\bibfnamefont {O.}~\bibnamefont {Vafek}},\ }\bibfield
   {title} {\enquote {\bibinfo {title} {{QED(3) theory of pairing pseudogap in
  cuprates. 1. From D wave superconductor to antiferromagnet via 'algebraic'
  Fermi liquid}},}\ }\href {\doibase 10.1103/PhysRevB.66.054535} {\bibfield
  {journal} {\bibinfo  {journal} {Phys. Rev.}\ }\textbf {\bibinfo {volume}
  {B66}},\ \bibinfo {pages} {054535} (\bibinfo {year} {2002})},\ \Eprint
  {http://arxiv.org/abs/cond-mat/0203333} {arXiv:cond-mat/0203333 [cond-mat]}
  \BibitemShut {NoStop}%
\bibitem [{\citenamefont {Laine}\ and\ \citenamefont
  {Vuorinen}(2016)}]{Laine:2016hma}%
  \BibitemOpen
  \bibfield  {author} {\bibinfo {author} {\bibfnamefont {Mikko}\ \bibnamefont
  {Laine}}\ and\ \bibinfo {author} {\bibfnamefont {Aleksi}\ \bibnamefont
  {Vuorinen}},\ }\bibfield  {title} {\enquote {\bibinfo {title} {{Basics of
  Thermal Field Theory}},}\ }\href {\doibase 10.1007/978-3-319-31933-9}
  {\bibfield  {journal} {\bibinfo  {journal} {Lect. Notes Phys.}\ }\textbf
  {\bibinfo {volume} {925}},\ \bibinfo {pages} {pp.1--281} (\bibinfo {year}
  {2016})},\ \Eprint {http://arxiv.org/abs/1701.01554} {arXiv:1701.01554}
  \BibitemShut {NoStop}%
\bibitem [{\citenamefont
  {Romatschke}(2019{\natexlab{c}})}]{Romatschke:2019rjk}%
  \BibitemOpen
  \bibfield  {author} {\bibinfo {author} {\bibfnamefont {Paul}\ \bibnamefont
  {Romatschke}},\ }\bibfield  {title} {\enquote {\bibinfo {title} {{Simple
  non-perturbative resummation schemes beyond mean-field: case study for scalar
  $\phi^4$ theory in 1+1 dimensions}},}\ }\href {\doibase
  10.1007/JHEP03(2019)149} {\bibfield  {journal} {\bibinfo  {journal} {JHEP}\
  }\textbf {\bibinfo {volume} {03}},\ \bibinfo {pages} {149} (\bibinfo {year}
  {2019}{\natexlab{c}})},\ \Eprint {http://arxiv.org/abs/1901.05483}
  {arXiv:1901.05483} \BibitemShut {NoStop}%
\bibitem [{\citenamefont {Romatschke}()}]{Romatschke:2019wxc}%
  \BibitemOpen
  \bibfield  {author} {\bibinfo {author} {\bibfnamefont {Paul}\ \bibnamefont
  {Romatschke}},\ }\bibfield  {title} {\enquote {\bibinfo {title} {{Simple
  non-perturbative resummation schemes beyond mean-field II: thermodynamics of
  scalar $\phi^4$ theory in 1+1 dimensions at arbitrary coupling}},}\
  }\href@noop {} {\ }\Eprint {http://arxiv.org/abs/1903.09661}
  {arXiv:1903.09661} \BibitemShut {NoStop}%
\bibitem [{\citenamefont {Kraemmer}\ and\ \citenamefont
  {Rebhan}(2004)}]{Kraemmer:2003gd}%
  \BibitemOpen
  \bibfield  {author} {\bibinfo {author} {\bibfnamefont {Ulrike}\ \bibnamefont
  {Kraemmer}}\ and\ \bibinfo {author} {\bibfnamefont {Anton}\ \bibnamefont
  {Rebhan}},\ }\bibfield  {title} {\enquote {\bibinfo {title} {{Advances in
  perturbative thermal field theory}},}\ }\href {\doibase
  10.1088/0034-4885/67/3/R05} {\bibfield  {journal} {\bibinfo  {journal} {Rept.
  Prog. Phys.}\ }\textbf {\bibinfo {volume} {67}},\ \bibinfo {pages} {351}
  (\bibinfo {year} {2004})},\ \Eprint {http://arxiv.org/abs/hep-ph/0310337}
  {arXiv:hep-ph/0310337 [hep-ph]} \BibitemShut {NoStop}%
\bibitem [{\citenamefont {Le~Bellac}(1996)}]{lebellac_1996}%
  \BibitemOpen
  \bibfield  {author} {\bibinfo {author} {\bibfnamefont {Michel}\ \bibnamefont
  {Le~Bellac}},\ }\href {\doibase 10.1017/CBO9780511721700} {\emph {\bibinfo
  {title} {Thermal Field Theory}}}\ (\bibinfo  {publisher} {Cambridge
  University Press},\ \bibinfo {year} {1996})\BibitemShut {NoStop}%
\bibitem [{\citenamefont {Carrington}\ and\ \citenamefont
  {Mrowczynski}(2019)}]{Carrington:2019ggz}%
  \BibitemOpen
  \bibfield  {author} {\bibinfo {author} {\bibfnamefont {Margaret~E.}\
  \bibnamefont {Carrington}}\ and\ \bibinfo {author} {\bibfnamefont
  {Stanislaw}\ \bibnamefont {Mrowczynski}},\ }\bibfield  {title} {\enquote
  {\bibinfo {title} {{Effective Coupling Constant of Plasmons}},}\ }\href@noop
  {} {\  (\bibinfo {year} {2019})},\ \Eprint {http://arxiv.org/abs/1907.03131}
  {arXiv:1907.03131 [hep-ph]} \BibitemShut {NoStop}%
\bibitem [{\citenamefont {Linde}(1980)}]{Linde:1980ts}%
  \BibitemOpen
  \bibfield  {author} {\bibinfo {author} {\bibfnamefont {Andrei~D.}\
  \bibnamefont {Linde}},\ }\bibfield  {title} {\enquote {\bibinfo {title}
  {{Infrared Problem in Thermodynamics of the Yang-Mills Gas}},}\ }\href
  {\doibase 10.1016/0370-2693(80)90769-8} {\bibfield  {journal} {\bibinfo
  {journal} {Phys. Lett.}\ }\textbf {\bibinfo {volume} {96B}},\ \bibinfo
  {pages} {289--292} (\bibinfo {year} {1980})}\BibitemShut {NoStop}%
\bibitem [{\citenamefont {Grozin}(2005)}]{Grozin:2005yg}%
  \BibitemOpen
  \bibfield  {author} {\bibinfo {author} {\bibfnamefont {Andrey}\ \bibnamefont
  {Grozin}},\ }\bibfield  {title} {\enquote {\bibinfo {title} {{Lectures on QED
  and QCD}},}\ }in\ \href@noop {} {\emph {\bibinfo {booktitle} {{3rd Dubna
  International Advanced School of Theoretical Physics Dubna, Russia, January
  29-February 6, 2005}}}}\ (\bibinfo {year} {2005})\ pp.\ \bibinfo {pages}
  {1--156},\ \Eprint {http://arxiv.org/abs/hep-ph/0508242}
  {arXiv:hep-ph/0508242 [hep-ph]} \BibitemShut {NoStop}%
\bibitem [{\citenamefont {Romatschke}(2019{\natexlab{d}})}]{codedown}%
  \BibitemOpen
  \bibfield  {author} {\bibinfo {author} {\bibfnamefont {P.}~\bibnamefont
  {Romatschke}},\ }\bibfield  {title} {\enquote {\bibinfo {title} {{Numerical
  codes for QED in 2+1 dimensions}},}\ }\href@noop {} {\bibfield  {journal}
  {\bibinfo  {journal} {https://github.com/paro8929/QED$\quad$}\ } (\bibinfo
  {year} {2019}{\natexlab{d}})}\BibitemShut {NoStop}%
\bibitem [{\citenamefont
  {Romatschke}(2019{\natexlab{e}})}]{Romatschke:2019mjm}%
  \BibitemOpen
  \bibfield  {author} {\bibinfo {author} {\bibfnamefont {Paul}\ \bibnamefont
  {Romatschke}},\ }\bibfield  {title} {\enquote {\bibinfo {title}
  {{Fractionalized Degrees of Freedom at Infinite Coupling in large Nf QED in
  2+1 dimensions}},}\ }\href@noop {} {\  (\bibinfo {year}
  {2019}{\natexlab{e}})},\ \Eprint {http://arxiv.org/abs/1908.02758}
  {arXiv:1908.02758 [hep-th]} \BibitemShut {NoStop}%
\end{thebibliography}%
\end{document}